\documentclass[%
 aip,
 jmp,%
 amsmath,amssymb,
reprint,%
]{revtex4-1}

\usepackage{graphicx}
\usepackage{dcolumn}
\usepackage{bm}
\usepackage{ulem}

\usepackage{color}
\usepackage{subfigure}
\usepackage[colorlinks,linkcolor=blue]{hyperref}

\begin{document}

\preprint{AIP/123-QED}

\title[{\color{red}Applied Physics Letters. 2019, 115(24), 244101~~~~~~~}{\href{https://doi.org/10.1063/1.5129778}{click to download- https://doi.org/10.1063/1.5129778}}]{Electric field measurements under DC corona discharges in ambient air by electric field induced second harmonic generation}
\author{Yingzhe Cui}
\author{Chijie Zhuang}
\thanks{Author to whom correspondence should be addressed. Electronic mail: chijie@tsinghua.edu.cn}
\author{Rong Zeng}
\affiliation{Department of Electrical Engineering, Tsinghua University, Beijing 100084, China}

\date{\today}

\begin{abstract}
Electric field distribution is critically important for quantitative insights into the physics of non-equilibrium plasma like corona. To analyze the electric field as well as the ion flow (space charge) distribution under DC corona discharges, the ion flow model has been widely adopted; Kaptzov's assumption, which states the {steady state} electric field at the conductor surface remains at the corona onset value, serves as a boundary condition. In this letter, we investigate the electric field distribution under DC corona discharges between coaxial cylindrical electrodes in ambient air by electric field induced second harmonic generation with nano-second pulse laser beams. The electric field distribution (with or without corona discharge) is obtained. By comparing the measurements with the results predicted by the ion flow model for negative corona discharge, it is found that the electric field at the conductor surface is proportional to the current density of the corona discharge with a negative constant of proportionality. Therefore, for negative corona discharges, Kaptzov's assumption is valid only when the discharge current approaches zero or is small.
\end{abstract}

\maketitle
A corona is a cold plasma that results from a locally enhanced electric field near an energized conductor or a metal tip, and has broad applications, such as the electrostatic precipitators \cite{item1}, polyester fabric\cite{item2}, polyaniline doping\cite{item3} and high voltage transmission line designs \cite{item4}. Therefore, regulating the corona discharge is greatly desired, especially for many engineering purposes. To better regulate the corona (as well as other non-equilibrium plasmas), the electric field distribution is critically important for quantitative insights because it is strongly correlated with the microphysical processes of discharges that involve electron transport, kinetics of ionization, electron energy partition, and generation of excited species and atoms\cite{item5,item6}.

However, when a self-sustained corona discharge emerges in a system, we are unable to calculate the electric field distribution by simply solving a Laplace equation. To analyze the spatial electric field distribution under corona discharges, the ion flow model, which accounts for most key features such as ion drift and recombination, has been widely adopted \cite{item7,du,zhang}. The model includes the transport equations and a Poisson equation.

To make the model complete, the value of the electric field at the surface of the electrode, which serves as the boundary condition of the model\cite{item7, du, zhang, item8}, is assumed to remain constant. The solution to the electric field and charge distributions under a corona discharge is sensitive to this value. Townsend first attempted to estimate this value and suggested that the electric field at the surface of the conductor remains at its onset value \cite{item9}. The basis of this assumption is that a reduction in value below the onset value results in the extinction of the discharge. Cobine suggested that a highly ionized region near a line conductor may require for a field greater than the onset field \cite{item10}. Nowadays, the assumption attributed to Kaptzov\cite{item7,du, zhang, item8} that the electric field at the surface of the conductor remains at the onset value, is extensively applied in both monopolar and bipolar ion flow analysis. 
However, few experimental results of the surface electric field are available. 

The lack of experimental results is attributed mainly to the difficulty of the electric field measurement during a discharge process. Much effort has been spent over the decades using intrusive methods, e.g., electrostatic fluxmeters \cite{item11} or electro-optical sensors based on the Pockels effect \cite{item12, item13}, all of which would inevitably disturb the electric field distribution. Non-intrusive methods appear more attractive in the sense that the field distribution remains undisturbed. Optical emission spectroscopy (OES) is useful; however, these emission-based methods are limited to places where the photons are emitted and do not work for regions outside the discharge area \cite{item14,item15,item16}. A four-wave mixing (FWM) technique does not have this limitation, and has been used to measure fields in H$_2$ and N$_2$ either with or without discharges; however, it mainly works in gases such as H$_2$ and N$_2$, and fails for atomic species \cite{item17,item18}. 

In summary, the electric field measurement in a discharge is critically important yet still difficult. 

Recently, a non-intrusive method, namely electric field induced second harmonic generation (E-FISHG)\cite{item19,item20,item21,item22}, has been used to measure the electric field in many cases like the surface plasma flow actuator \cite{item23} and fast ionization wave \cite{item24}. Unlike OES and FWM, the E-FISHG technique is not species-dependent, and is more sensitive than FWM.

In this letter, we measured the electric field under negative DC corona discharges in ambient air using E-FISHG with nano-second pulse laser beams, and studied the validity of Kaptzov's widely adopted  assumption.

The E-FISHG is a third-order nonlinear process involving the laser field and external field \cite{item19,item20,item21,item22}. The induced polarization at the second harmonic frequency $2\omega$, denoted by $P_i^{(2\omega)}$, is calculated from\cite{item19,item20,item21}
\begin{equation}
P_i^{(2\omega)}=\frac{3}{2}N \chi_{i,j,k,l}^{(3)}(-2\omega, 0, \omega, \omega)E_j^F E_k^\omega E_l^\omega,
\end{equation}
where $E_{k,l}^\omega$ denotes the electric field of the incedient laser which are the same in second-harmonic generation; $E_j^F$ is the electric field to be measured; $N$ is the number density of the molecular gas; and $\chi_{i,j,k,l}^{(3)}$ is the nonlinear susceptibility tensor depending on the molecular dipole moments and field orientations \cite{item25}, and the subscripts of the susceptibility tensor denote to the polarization of the second harmonic beam, the electric field, and the first and second laser beams, respectively.

With the plane-wave approximation, the electric field obtains from the relation between the intensities of the pump laser $I^\omega$ and the induced second harmonic signals $I^{2\omega}$\cite{item20},
\begin{equation}
I^{2\omega}\propto N^2 (E_{\mbox{ext}})^2 (I^\omega)^2, 
\label{iroot}
\end{equation} 
where $E_{\mbox{ext}}$ denotes the external electric field.

Electric field measurements were conducted under DC corona discharges in ambient air between coaxial stainless steel cylindrical electrodes. The setup of the nanosecond E-FISHG measurement system (shown in Fig. 1(a)) includes a 1064 nm Nd:YAG laser producing an output beam with a pulse duration of approximately 5 ns and pulse energy of 60 mJ. The beam first passes through a dispersion prism (DP), and is focused at the center of the discharge section (HV) using an f=300 mm plano-convex lens (FL1). A long pass filter (LP) is placed before the discharge section to remove stray second harmonic signals. An electric voltage is applied across the discharge section. 
A second harmonic signal is generated collinearly with the pump beam, and the beam is recollimated using an f=150 mm plano-convex lens (FL2). The fundamental beam is then separated from the second harmonic beam by a dichroic mirror (M1) and monitored by a PIN photodiode (PD). The second harmonic signal is reflected by another dichroic mirror (DM) and focused into a monochromator with a resolution higher than 0.1 nm, by an f=250 mm plano-convex lens (FL3), and then detected by a photomultiplier (PMT; Hamamatsu Photonics) which is installed in a metal cylinder to reduce electromagnetic interference. A narrowband pass filter (BF, 532 nm, FWHM 10 nm) is placed at the entrance of the PMT to remove stray light. The PMT and photodiode signals are recorded by an oscilloscope (DPO4054, Tektronix; 2.5 GHz sampling rate, 500 MHz bandwidth). 

For the electrode configuration (Figs. 1(b) and 1(c)), the inner electrode is a conductor with a radius of $r_0=0.1$ cm, and the outer electrode is a 4.4-cm-long cylinder of inner radius $R=3$ cm. The electrodes were polished using waterproof abrasive paper and cleaned with alcohol before the experiments. All the electrodes were fixed to an insulated platform. To measure the electric field at different positions, a spiral ruler (minimum spatial resolution 10~$\mu$m) is used to translate the insulated platform along the x-axis. 

The diameter of the laser beam at the focal point was estimated to be 210 $\mu$m, and the Rayleigh range was about 3 cm. The electric field is along the radial direction due to the axial symmetry of the experiment setup. To increase the signal to noise ratio, the laser beam is polarized parallel to the electric field \cite{item19,item22}.

During the experiments, the inner conductor was grounded and a DC voltage was applied to the outer cylinder electrode; the corona current was then recorded. 
The length of the electrode is {approximately $50\%$ larger than the distance between the two electrodes, and we performed a numerical analysis for the electric field distribution which shows that except the outermost 0.5 cm near the ends along the electrodes, the electric field distributes relatively uniformly, e.g., the maximal and minimum electric fields differ 3.7\% and 1.6\% from the root mean square (RMS) average of the electric field at $r=0.25$ cm, respectively; and differ 1.7\% and 3.4\% from the RMS average at $r=1$ cm, respectively.}

Because the emission spectrum of an atmospheric corona discharge mainly ranges between 200 and 405 nm which is shorter than 532 nm\cite{item26}, the background harmonic noise emitted from the corona discharges is not considered. 

\begin{figure}[!htbp]
\centering
\includegraphics[width=8.5 cm]{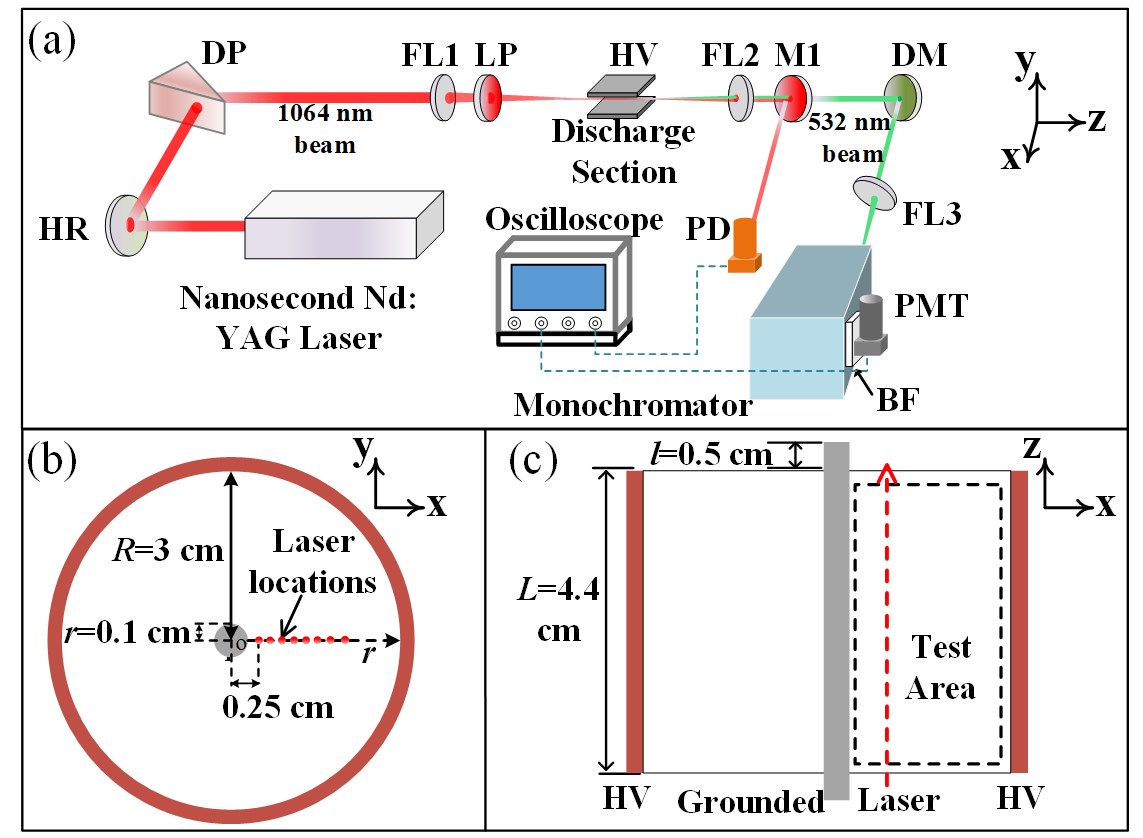}
\caption{\label{fig1} Schematic of the corona discharge experiment to measure the electric field between a pair of coaxial cylindrical electrodes using nanosecond second harmonic generation: (a) Setup for E-FISH measurements. HR: a 1064-nm reflector; DP: dispersion prism; FL1: 300 mm plano-convex lens; FL2: 150 mm plano-convex lens; FL3: 250 mm plano-convex lens; LP: long pass filter; M1: dichroic mirror (1064 nm reflecting, 532 nm transmitting); DM: dichroic mirror (532 nm reflecting, 1064 nm transmitting); PD: photodiode; PMT: photomultiplier tube; BF: 532 nm bandpass filter. (b) Axial view and (c) top view of the electrodes. }
\end{figure}

Before conducting the experiments, we calibrated the system. Without any corona discharges, the electric field (which we hereafter refer to as the Laplacian field) at each spatial location $r$ in the area between the coaxial cylindrical electrodes is
\begin{equation}
E(r) = \frac{U}{r\ln(R/r_0)}, \label{lap}
\end{equation} 
where $U$ is the voltage between the two electrodes. 

We applied different DC voltages, and monitored the associated pump beam laser intensity and second harmonic intensity response to the laser shots. Note that the pump beam intensity is determined by integrating the waveform obtained by the photodiode over the full response time, and the second harmonic intensity is determined in the same way \cite{item20}. Fig. 2 shows an example of the relationship between the square root of the second harmonic intensity and the electric field obtained at $r=0.25$ cm calculated using Eq. (\ref{lap}). Each datum was the average value from 30 signals. As indicated by Eq. (\ref{iroot}), the square root of the second harmonic signal is proportional to the applied electric field.

\begin{figure}[!htbp]
\centering
\includegraphics{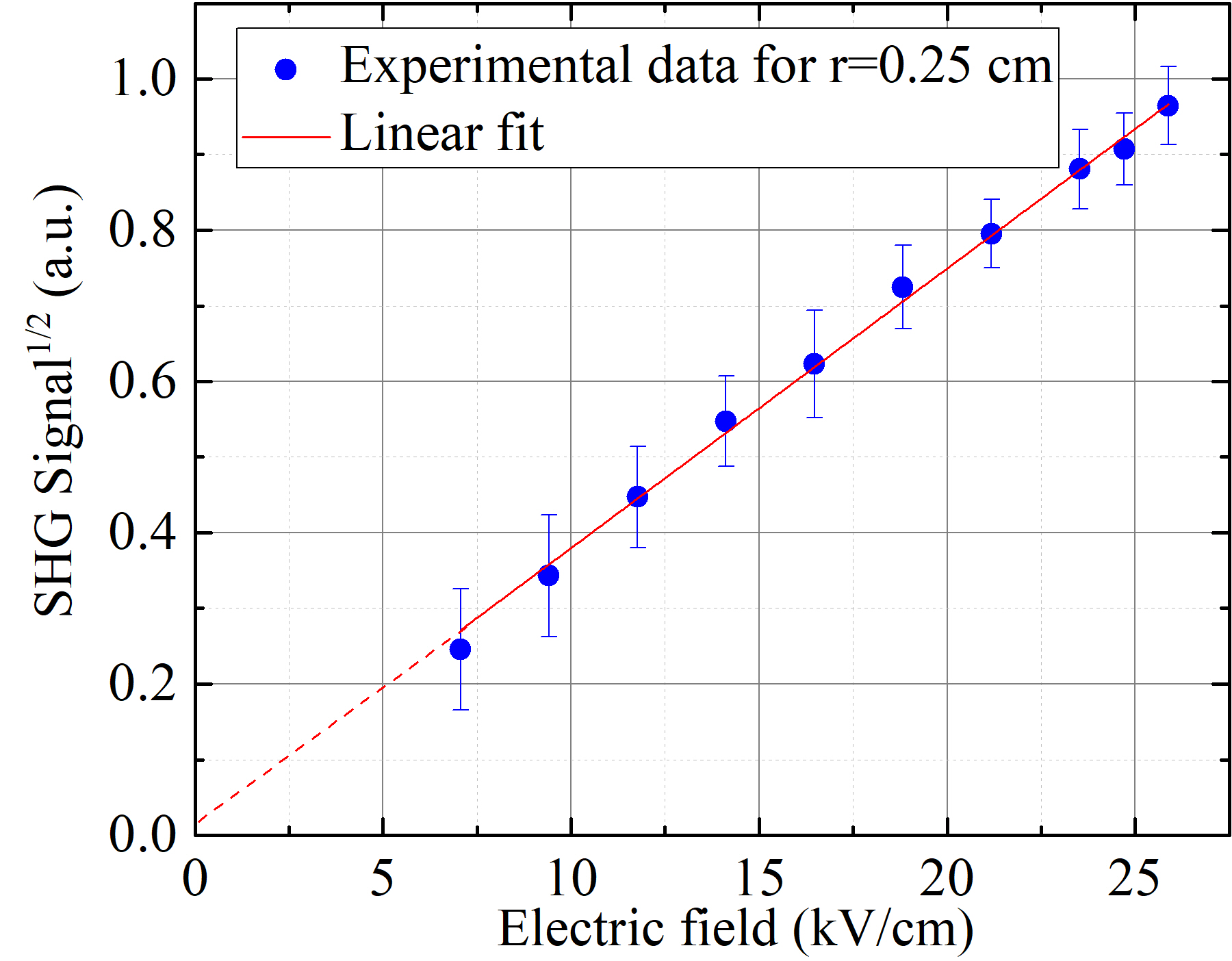}
\caption{\label{fig2}Calibration of the electric field at $r=0.25$ cm between the coaxial cylindrical electrodes. Single-shot data were collected, each datum point being the average value from 30 signals. }
\end{figure} 

Next, we measured the electric field distribution between the electrodes under different DC voltages. Figure 3 shows the results of the coaxial cylindrical electrodes with an applied voltage of -20 kV, without any corona discharge. Each measured datum is the average value of 50 signals. The electric field determined by the E-FISHG method matches closely the theoretical values. Noting that the laser beam was unable to be infinitely close to the conductor surface (minimal distance was 1.5~mm in this study), otherwise optical noise would be generated when the laser hit the plastic bracket supporting the inner conductor electrode.

\begin{figure}[!h]
\centering
\includegraphics{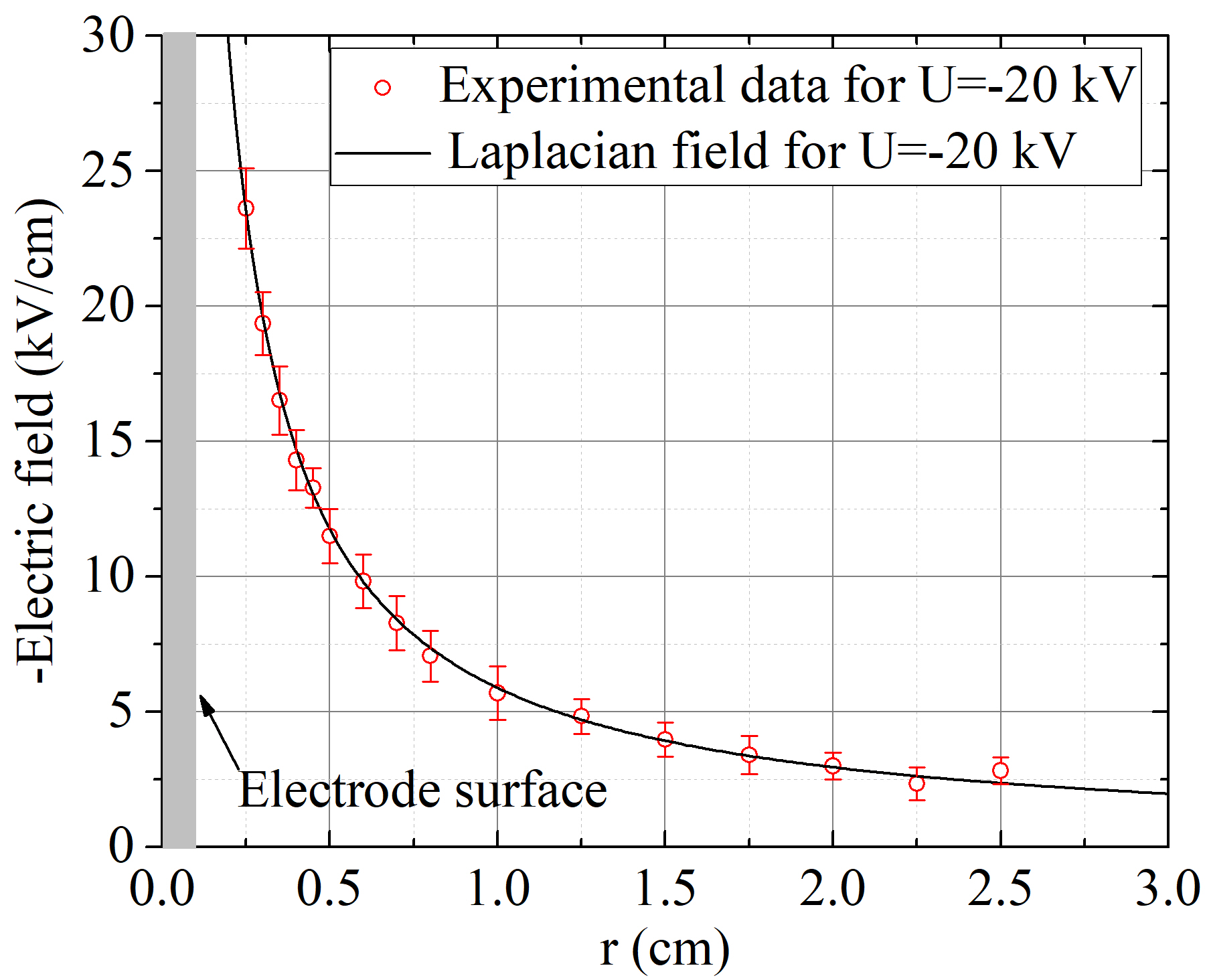}
\caption{\label{fig3} Electric field distribution of the coaxial cylindrical electrodes without any corona plasma. The applied DC voltage was -20 kV. Red circles: experimental results using nanosecond E-FISHG method, average value from 50 signals. }
\end{figure}

Figure 4 shows the spatial resolved  electric field measurements under different applied voltages above the corona onset threshold,  differing noticeably from the Laplacian field determined from Eq. (\ref{lap}). As the applied voltage and hence the corona discharge current increases, the difference between the measured and Laplacian fields becomes more evident. In addition,  from the distortion of the electric field distribution, the space charge was estimated to be of order $10^{-4}$ $\mu$C/cm$^3$. 

{It should be noted that the measurements are actually of a steady state solution for the electric field. Each datum is the average value of 50 signals, i.e., we randomly sampled the signals over the time domain, and the mean values of the electric field over the time are expected to be obtained after averaging. If we integrate the electric field along the axial direction ($U=\int E\mbox{d}r$), the integral agrees with the applied DC voltage well; For Fig. 4(a)-(c), the relative differences are 2.8\%, 3.9\% and 0.8\%, respectively.}

\begin{figure}[!htbp]
\centering
\includegraphics{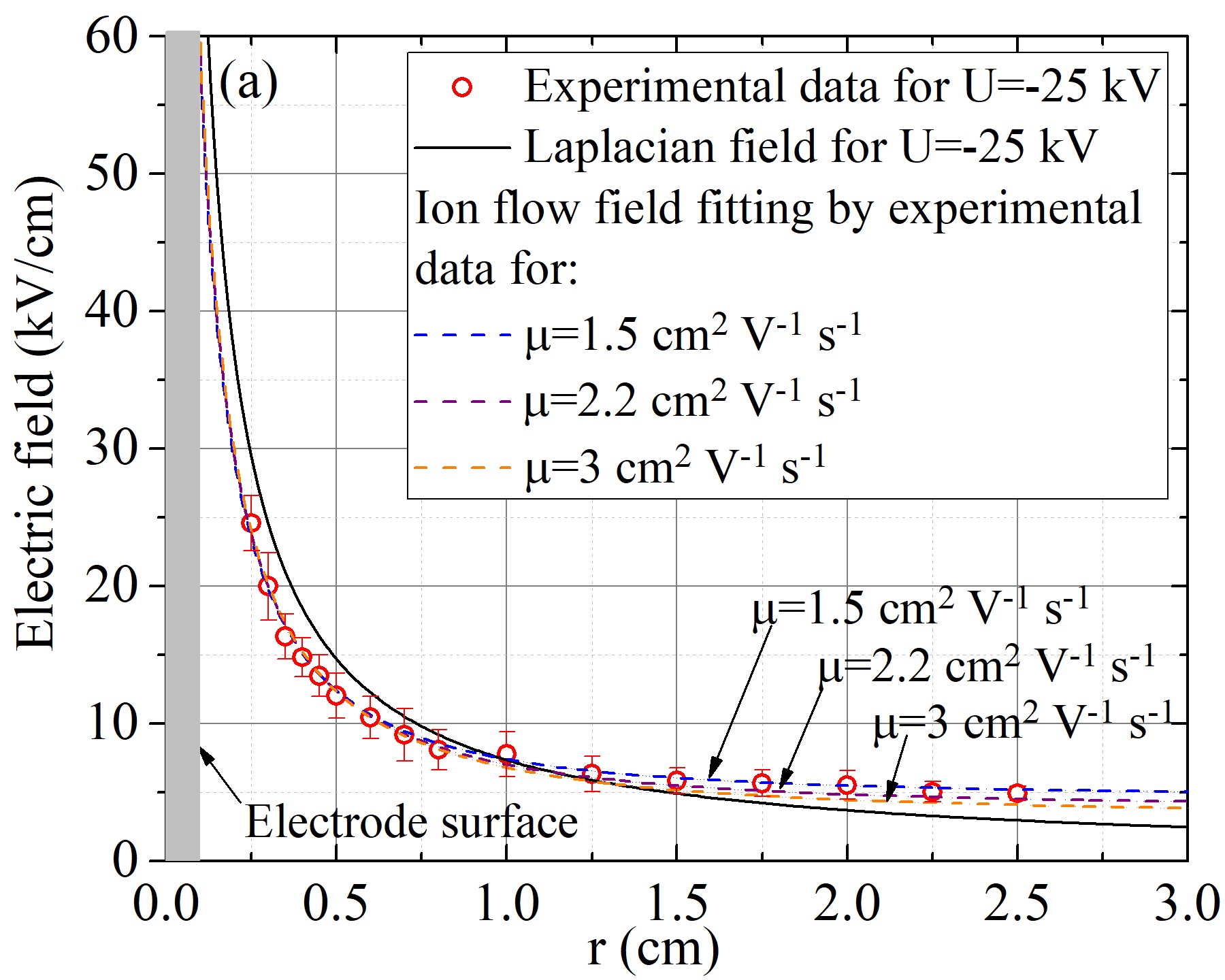}
\includegraphics{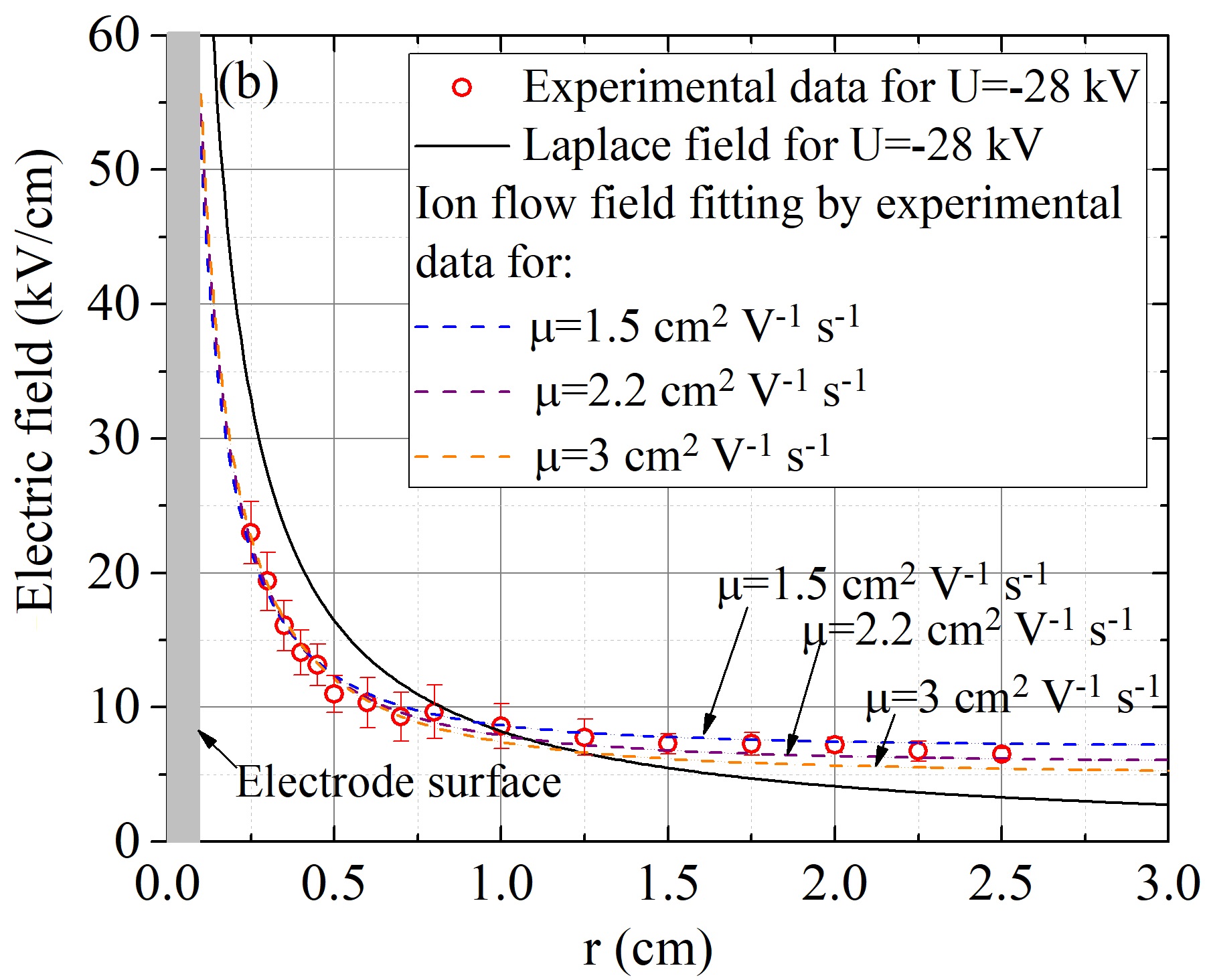}
\includegraphics{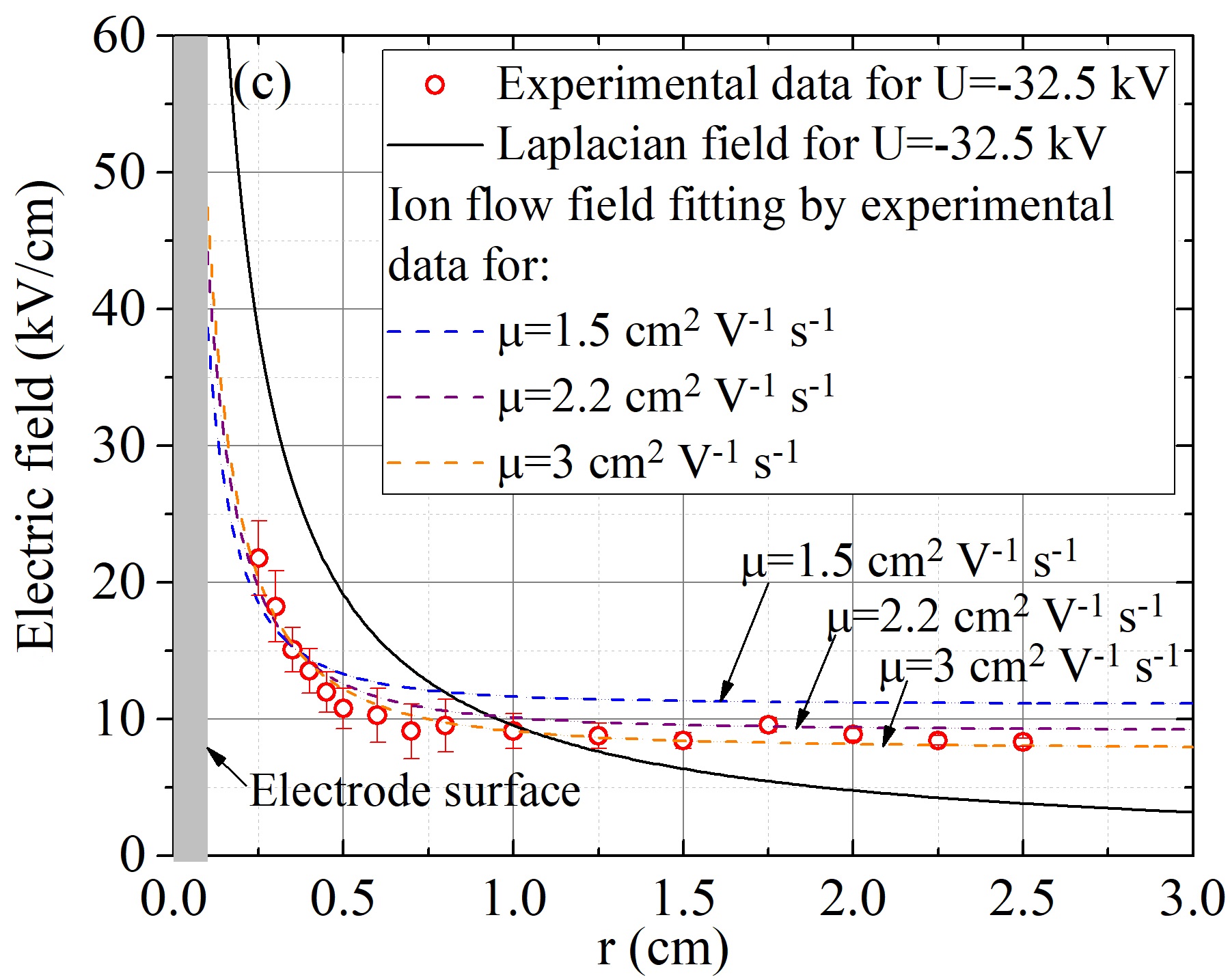}
\caption{\label{fig4} Electric field distribution between the coaxial cylindrical electrodes with corona plasma. The applied DC voltage and measured corona current are (a) $U=-25$ kV, $I=0.08$ mA; (b) $U=-28$ kV, $I=0.18$ mA; (c) $U=-32.5$ kV, $I=0.45$ mA, respectively. Red circles represent experimental results obtained using the nanosecond E-FISHG method and averaged over 50 signals.}
\end{figure}

Under the coaxial cylindrical configurations, the ion flow under a negative DC corona can be described by the continuity and Poisson equations, as Eq. (\ref{model}) \cite{item7,item29},
\begin{eqnarray}
I_l = 2\pi r \rho_r \mu E_r, ~~~~ \frac{1}{r}\frac{d}{dr}(rE_r)=\frac{\rho_r}{\varepsilon_0}, \label{model}
\end{eqnarray} 
where $r$ denotes the radial position; $I_l$ is the corona current density per unit length; $\rho_r$ is the spatial net charge density; $\mu$ is the mobility of negative ions; $r_0$ is the conductor radius. Assuming that $E_0$ is the electric field at the surface of the conductor, one obtains from Eq. (\ref{model}),
\begin{equation}
E(r)= \sqrt{\frac{I_l}{2\pi\varepsilon_0\mu}(1-\frac{r_0^2}{r^2})+(\frac{E_0r_0}{r})^2}. \label{solu}
\end{equation}

Note that the mobility of negative ions, which is assumed to be constant, may here weakly depend on the electric field, and typically varies from 1.5 to 3.0 cm$^2$/(V$\cdot$s) in ambient air \cite{item29}. We fitted the measured electric field distributions according to Eq. (\ref{solu}) with three values of $\mu$ (1.5, 2.2 and 3.0 cm$^2$/(V$\cdot$s)) to study the influence of the mobility of negative ions; the results are also plotted in Fig.\ref{fig4}. The fitting curves and measured data are in reasonable agreement. The difference between different values of $\mu$ are acceptable, especially when the position nears the conductor surface. We therefore assume $\mu$=2.2 cm$^2$/(V$\cdot$s) in the following analysis. 

As already mentioned above, the laser beam cannot be infinitely close to the conductor surface. We may alternatively obtain the electric field at the surface ($E_0$) by numerically fitting the measured electric field data using formula (5). To achieve this {objective}, more measurements of electric field distributions besides the data shown in Fig. 4 were taken under different corona currents.

Figure 5 shows the electric field at the conductor surface with different corona current densities with an assumed ion mobility of 2.2 cm$^2$/(V$\cdot$s). The error bars in Fig. 5 represent the 95\% confidence interval when determining the surface electric field. 

\begin{figure}[!htbp]
\centering
\includegraphics{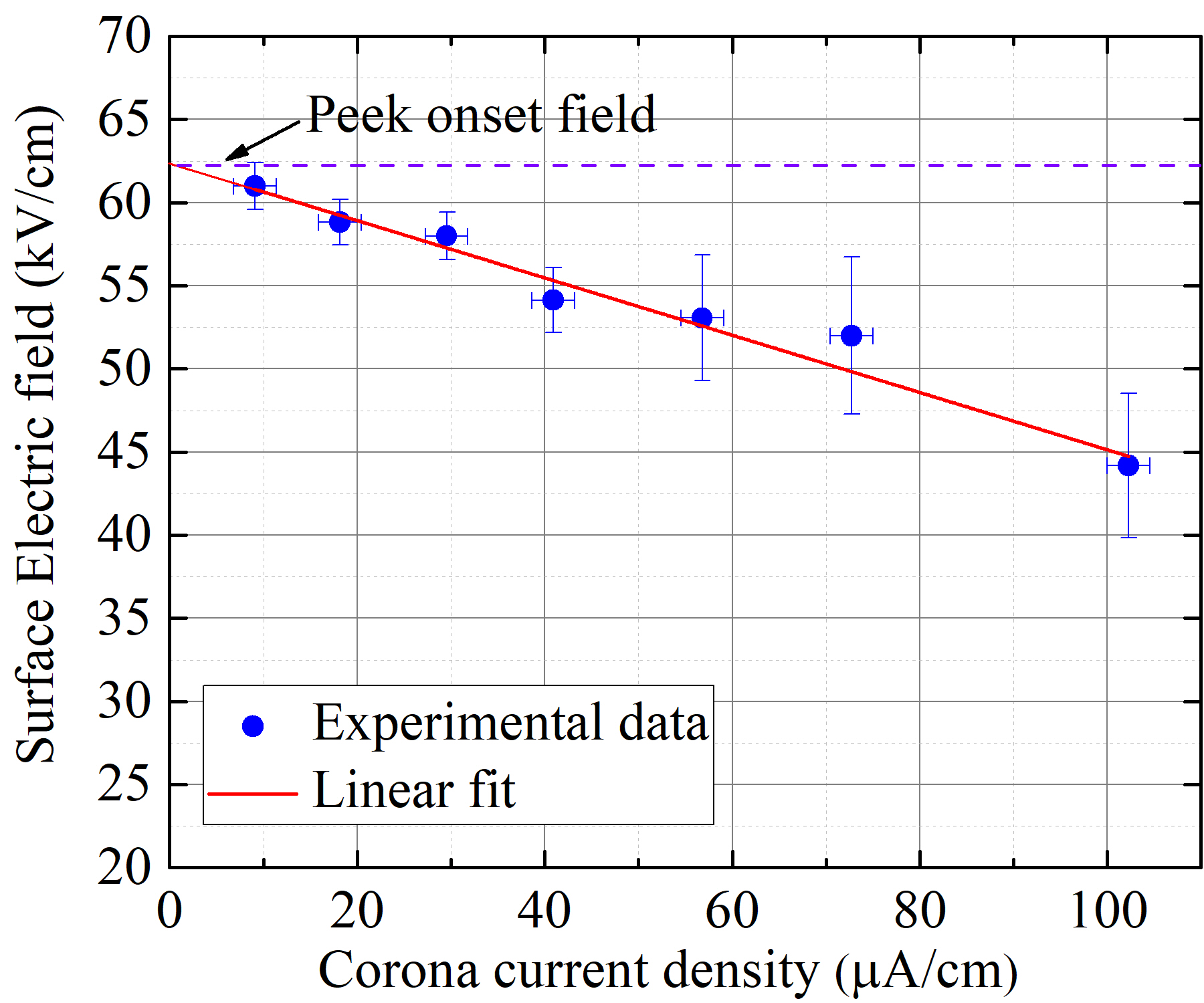}
\caption{\label{fig5} Dependence of surface electric field of a negative corona on corona current.}
\end{figure}

Contrary to Kaptzov's assumption, Fig. \ref{fig5} shows that the electric field at the conductor surface is roughly proportional to the corona current density with a negative constant of proportionality. According to Kaptzov’s assumption, the electric field at the conductor's surface remains constant at the value determined using Peek's formula \cite{item30}: 
\begin{equation}
E_0 = E(r=r_0)=31 \delta (1+\frac{0.308}{(\delta r_0)^{\frac{1}{2}}})
\end{equation} 
where the relative air density $\delta$ is assumed to be 1.0 here. The surface field approaches the corona onset field determined by Peek's formula only when the corona current approaches zero. 

We also studied the influence of the ion mobility on determining surface electric field. For $U=-25,-32.5$ kV, the variation of the determined surface field is no more than 3\% and 6\%,respectively, when the ion mobility varies from 1.5 to 3 cm$^2$/(V$\cdot$s).

{In literatures, some theoretical studies such as [\onlinecite{item31}] predicted that the surface field was only slightly changed for corona currents up to 100 $\mu$A/cm; Waters gave a result that the surface electric field decreases as the current density increases for negative coronas\cite{item11}; Meanwhile, Zheng predicted a decrease of surface electric field according to the voltage-current characteristics for negative coronas \cite{zheng2}. As the applied voltage and hence the corona discharge current increases, the Laplacian electric field at the conductor surface increases; meanwhile, the existence of negative ions in space could weaken the electric field at the surface of the conductor which is with a negative voltage; therefore, the reason for the variation of the overall electric field in steady state considering both effects requires a more detailed investigation.}

In summary, we measured the {steady state} electric field under negative DC corona discharges in ambient air by E-FISHG with nano-second pulse laser beams. By comparing the measurements with the results predicted by the ion flow model for corona discharge, we show that for negative corona discharges, Kaptzov's widely adopted assumption is valid only when the discharge current density approaches zero or is small.

~\\
This work is supported by National Natural Science Foundation of China under grant 51577098.

\nocite{*}
%
\end{document}